\documentclass[apjl]{emulateapj}
\usepackage[colorlinks=true,citecolor=blue,urlcolor=red,linkcolor=blue]{hyperref}

\def\CII{\hbox{[C\,{\sc ii}]157.7$\,\mu$m}}
\def\CIIno{\hbox{[C\,{\sc ii}]}}

\def\H2{\hbox{H$_{2}$}}

\def\deg{$^{\circ}$}

\def\kms{${\rm km~s}^{-1}$}

\def\Lsun{\hbox{$L_\odot$}}

\def\LIR{\hbox{$L_{\rm IR}$}}
\def\LFIR{\hbox{$L_{\rm FIR}$}}

\def\G0{\hbox{$G_{\rm 0}$}}

\def\kms{\hbox{km$\,$s$^{-1}$}}

\def\Hubble{\hbox{km$\,$s$^{-1}\,$Mpc$^{-1}$}}

\shorttitle{The Strikingly Uniform, Highly Turbulent ISM of W2246-0526}
\shortauthors{D\'iaz-Santos et al.}

\begin{document}

\title{The Strikingly Uniform, Highly Turbulent Interstellar Medium of the Most Luminous Galaxy in the Universe}


\author{T.~D\'{\i}az-Santos\altaffilmark{1,*},
R.~J.~Assef\altaffilmark{1},
A.~W.~Blain\altaffilmark{2},
C.-W.~Tsai\altaffilmark{3},
M.~Aravena\altaffilmark{1},
P.~Eisenhardt\altaffilmark{3},
J.~Wu\altaffilmark{4},
D.~Stern\altaffilmark{3}
\&
C.~Bridge\altaffilmark{3}
}

\altaffiltext{*}{Contact email: tanio.diaz@mail.udp.cl}
\affil{$^{1}$N\'ucleo de Astronom\'ia de la Facultad de Ingenier\'ia, Universidad Diego Portales, Av. Ej\'ercito Libertador 441, Santiago, Chile}
\affil{$^{2}$University of Leicester, Physics and Astronomy, University Road, Leicester LE1 7RH, UK}
\affil{$^{3}$Jet Propulsion Laboratory, California Institute of Technology, 4800 Oak Grove Dr., Pasadena, CA 91109, USA}
\affil{$^{4}$Division of Physics \& Astronomy, University of California Los Angeles, Physics and Astronomy Building, 430 Portola Plaza, Los Angeles, CA 90095-1547, USA}

\begin{abstract}

Observed at \textit{z}\,=\,4.601 and with $L_{\rm bol}$\,=\,3.5$\,\times\,10^{14}\,\Lsun$, W2246-0526 is the most luminous galaxy known in the Universe, and hosts a deeply-buried active galactic nucleus (AGN)/super-massive black hole (SMBH). Discovered using the \textit{Wide-field Infrared Survey Explorer} (WISE), W2246-0526 is classified as a Hot Dust Obscured Galaxy (Hot DOG), based on its luminosity and dust temperature. Here we present spatially resolved ALMA \CII\, observations of W2246-0526, providing unique insight into the kinematics of its interstellar medium (ISM). The measured \CIIno\,-to-far-infrared ratio is $\sim\,2\,\times\,10^{-4}$, implying ISM conditions that compare only with the most obscured, compact starbursts and AGN in the local Universe today. The spatially resolved \CIIno\, line is strikingly uniform and very broad, 500--600\,\kms\, wide, extending throughout the entire galaxy over about 2.5\,kpc, with modest shear. Such a large, homogeneous velocity dispersion indicates a highly turbulent medium. W2246-0526 is unstable in terms of the energy and momentum that are being injected into the ISM, strongly suggesting that the gas is being blown away from the system isotropically, likely reflecting a cathartic state on its road to becoming an un-obscured quasar. W2246-0526 provides an extraordinary laboratory to study and model the properties and kinematics of gas in an extreme environment under strong feedback, at a time when the Universe was 1/10 of its current age: a system pushing the limits that can be reached during galaxy formation.

\end{abstract}


%________________________________________________________________
\section{Introduction}\label{s:intro}

Studying the most luminous galaxies in the Universe is a key step to understanding how the most massive galaxies assemble and evolve \citep[e.g.,][]{Hopkins2008I} -- all the way from their first appearance, about 500 million years after the Big Bang \citep{Zheng2012}, until today. Among the different populations of luminous galaxies currently known, Hot Dust-Obscured Galaxies (Hot DOGs; \citealt{Eisenhardt2012}), discovered by NASA's \textit{Wide-field Infrared Survey Explorer} (WISE; \citealt{Wright2010}), are the most extreme in terms of their luminosities and unusual hot dust temperatures \citep{Wu2012,Tsai2015}. The infrared (IR) emission from Hot DOGs is dominated by obscured accretion onto a central super-massive black hole (SMBH), in most cases without significant contribution from star formation \citep{Jones2014,Wu2014,Assef2015}. The large contrast between the underlying host galaxy and the hyper-luminous emission from the active galactic nucleus (AGN) implies that either the SMBH is much more massive than expected for the stellar mass of its host, or is radiating well above its Eddington limit \citep{Assef2015,Tsai2015}. Detailed analyses of their spectral energy distributions (SEDs) show that the rest-frame UV--near-IR emission of most Hot DOGs is consistent with a typically star-forming, not strongly obscured underlying host galaxy. Such luminous AGN, hosted by otherwise normal galaxies are likely to be at a key stage of their evolution --when feedback from the AGN may have started quenching star formation--, just about to become regular quasars and decay into dead elliptical galaxies \citep{Sanders1988a}.
% -- a stage that is predicted by theoretical models of galaxy evolution \citep{Hopkins2008I}. Hot DOGs like W2246-0526 are . 
W2246-0526 is the most extreme of these remarkable systems known. At a redshift \textit{z}\,=\,4.601\,$\pm$\,0.001, it is the most luminous galaxy known in the Universe \citep{Tsai2015}. 

Tracing the spatially resolved far-IR (FIR) dust emission and the kinematics of the interstellar medium (ISM) provides unique insight into how extreme AGN feedback may be operating. In normal star-forming galaxies and starbursts, \CIIno\, emission is most likely produced in photo-dissociation regions (PDRs; see, e.g., \citealt{Stacey2010,DS2013,Carilli2013}) associated with a source of intense far-UV (FUV) flux. On the other hand, it is not expected to be significantly powered by direct AGN photo-ionization, as this would progressively increase the ionization state of carbon beyond C$^{+}$ \citep{Langer2015}. Furthermore, unless there is considerable shielding by dust grains \citep[e.g.,][]{DS2010a,Esquej2014}, polycyclic aromatic hydrocarbon (PAH) molecules can be photo-dissociated within the central kpc of a powerful AGN \citep{Voit1992}. However, the \CIIno\,/PAH ratio is roughly independent of FIR luminosity and dust temperature in galaxies, both with and without an AGN (\citealt{Helou2001}, Armus et al., in prep.), thus suggesting that AGN do not contribute significantly to the direct heating of C$^+$ ions. It is possible, however, that turbulent heating related to the injection of mechanical energy due to shocks produced by AGN jets or supernovae winds may contribute substantially to \CIIno\, emission \citep{Meijerink2013,Appleton2013,Guillard2015}.

In this letter we present spatially-resolved \CIIno\, observations obtained with ALMA of the exceptional ISM in W2246-0526, the most luminous galaxy known in the Universe. Throughout the paper, we have used a cosmology with the following parameters: $\Omega_{\rm M}$\,=\,0.3, $\Omega_{\Lambda}$\,=\,0.7 and $H_0$\,=\,70\,$\Hubble$.

\section{Observations and Analysis}\label{s:obs}

As part of an ALMA cycle-2 campaign (ID: 2013.1.00576.S; PI: R. Assef) aimed at studying the ionized gas in the most luminous Hot DOGs, we were awarded 8.1h in bands 7 and 8 to study the red-shifted 157.7$\mu$m ($^2P_{3/2}\,\rightarrow\,^2P_{1/2}$) fine-structure transition of ionized carbon, \CIIno, and the underlying dust continuum emission in a sample of Hot DOGs spanning a range of redshifts and luminosities. Observations of W2246-0526, the first galaxy of the sample observed, were carried out on 2014-06-29 and 2014-12-07 in band 7 using four spectral windows in the largest bandwidth mode (1.875\,GHz). The \CIIno\, line was placed at the center of the reference spectral window (spw0), assuming the redshift derived from optical lines \citep[$z$\,=\,4.593;][]{Wu2012}; the other three spectral windows probing the continuum blue-ward of the line. The on-source integration time was 956\,s in array configuration C34-4. Flux, phase, and band-pass calibrators were also obtained, for a total time of 3330\,s. 

We used the CASA software to process and clean the data. The cleaning algorithm was run using a natural weighting for the u-v visibility plane\footnote{A uniform weighting does not provide further spatial information on the source due to the rapid decrease of S/N.}. The angular resolution of the observations (clean beam FWHM) is 0.386\arcsec\,$\times$\,0.356\arcsec, with a major-axis position angle (P.A.) of 47\deg\, (see Figure~\ref{f:images}). The r.m.s. of the observations is 0.6 mJy/beam, measured in 35\,\kms\, channel bins. The \CIIno\, emission line ($z_{\rm [CII]}\,=\,4.601\,\pm\,0.001$) was observed in a spectral window that missed part of the red wing of the line, due to a significant blue-shift of the UV/optical emission lines as compared with the ISM, a difference that is common in high-redshift dusty galaxies and quasars.

A Gaussian fits the emission profile well, and is consistent with only 7\% of the total line flux lying outside of the observed window (see Figure~\ref{f:spectra}, left). The total Gaussian flux measured from a spectrum extracted with a 1\arcsec\, diameter circular aperture centered on W2246-0526 is 9.3\,$\pm$\,0.3\,Jy\,\kms. The observed-frame continuum flux density underlying the emission line is 7.4\,$\pm$\,0.6\,mJy. Using an asymmetric, 2-D Gaussian function to fit the observed widths of the \CIIno\, line and dust continuum emission we find them to be 0.539\arcsec\,$\times$\,0.511\arcsec\, (P.A.\,=\,122\deg), and 0.433\arcsec\,$\times$\,0.407\arcsec\, (P.A.\,=\,33\deg), respectively.

\section{Results}\label{s:results}

\begin{figure}[t!]
\vspace{.25cm}
\epsscale{1.15}
\plotone{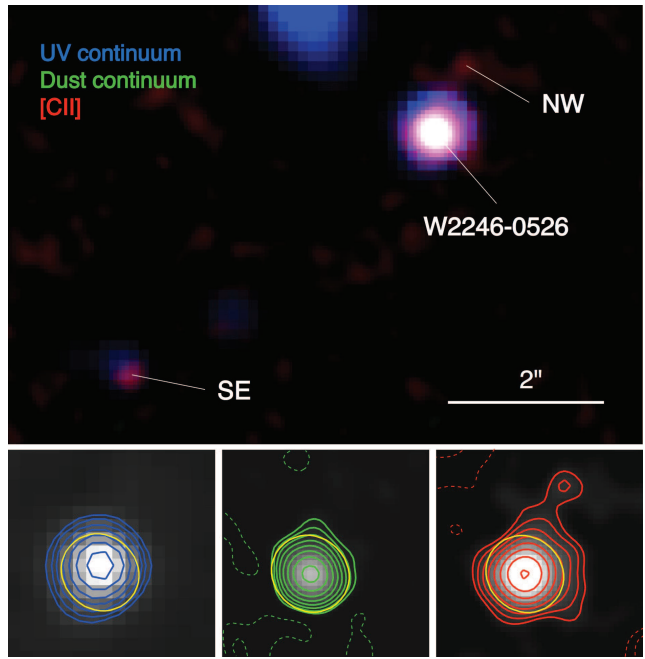}
\vspace{.25cm}
\caption{\footnotesize Multi-waveband ALMA and \textit{HST} image of W2246-0526 (top panel), including serendipitous north-west (NW) and south-east (SE) companions, located 6.6 and 33\,kpc away from W2246-0526 respectively. ALMA \CIIno\, line emission is shown in red, dust continuum emission is in shown in green, and \textit{HST} rest-frame UV emission (2860\AA) is shown in blue. The \textit{HST} image has been smoothed to match the ALMA resolution. The lower panels show the emission at each wavelength. Solid contour levels step in units of 3$\,\sigma\,$ times 2$^{n/2}$; while dashed negative contours are at $-1$ and $-2\,\sigma$. The clean ALMA half-maximum power beam is represented by the yellow contour. 1\arcsec\,=\,6.7 kpc in W2246-0526.}\label{f:images}
\vspace{.25cm}
\end{figure}

\begin{figure*}[t!]
\vspace{.25cm}
\epsscale{.38}
\plotone{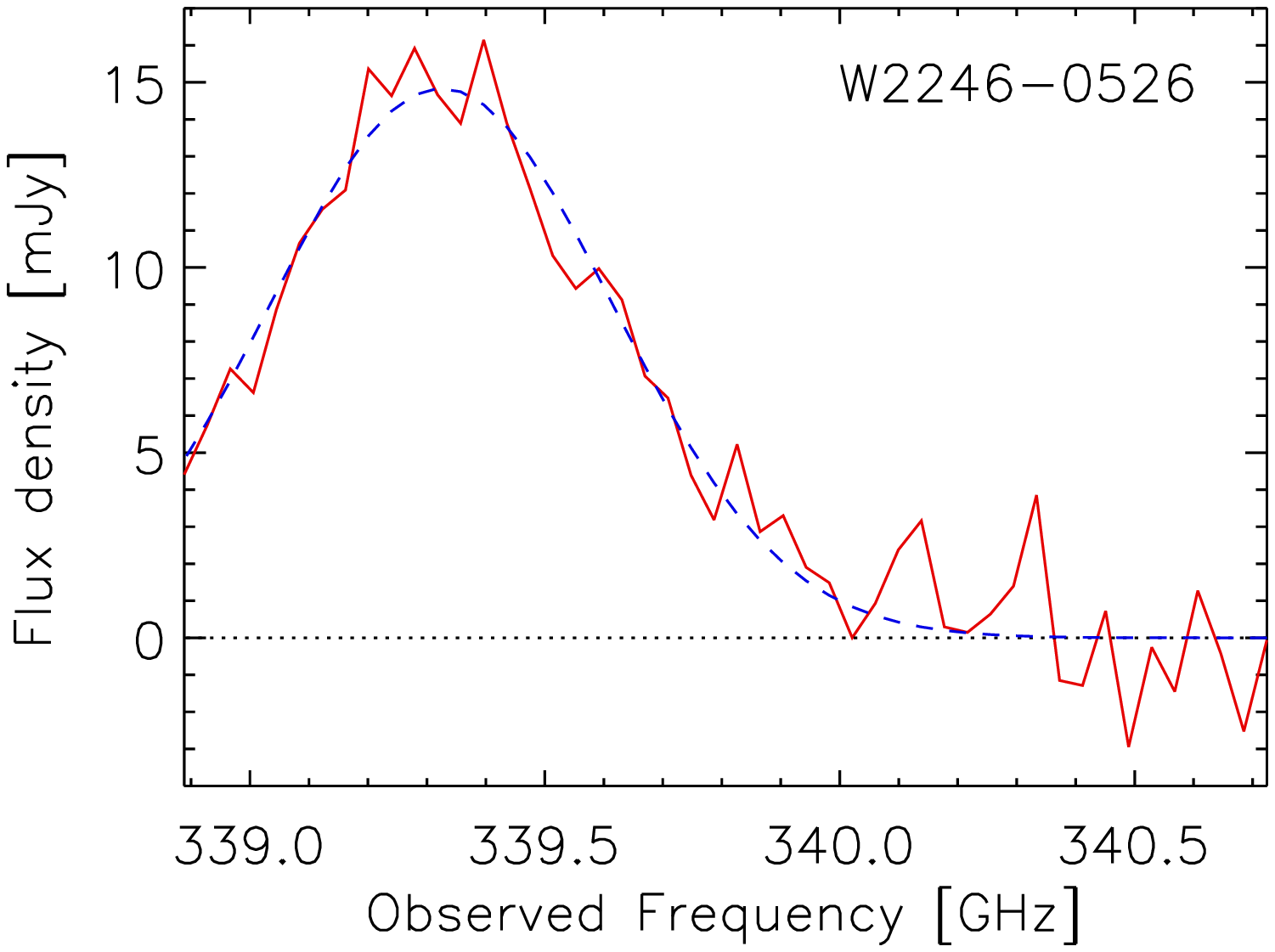}\plotone{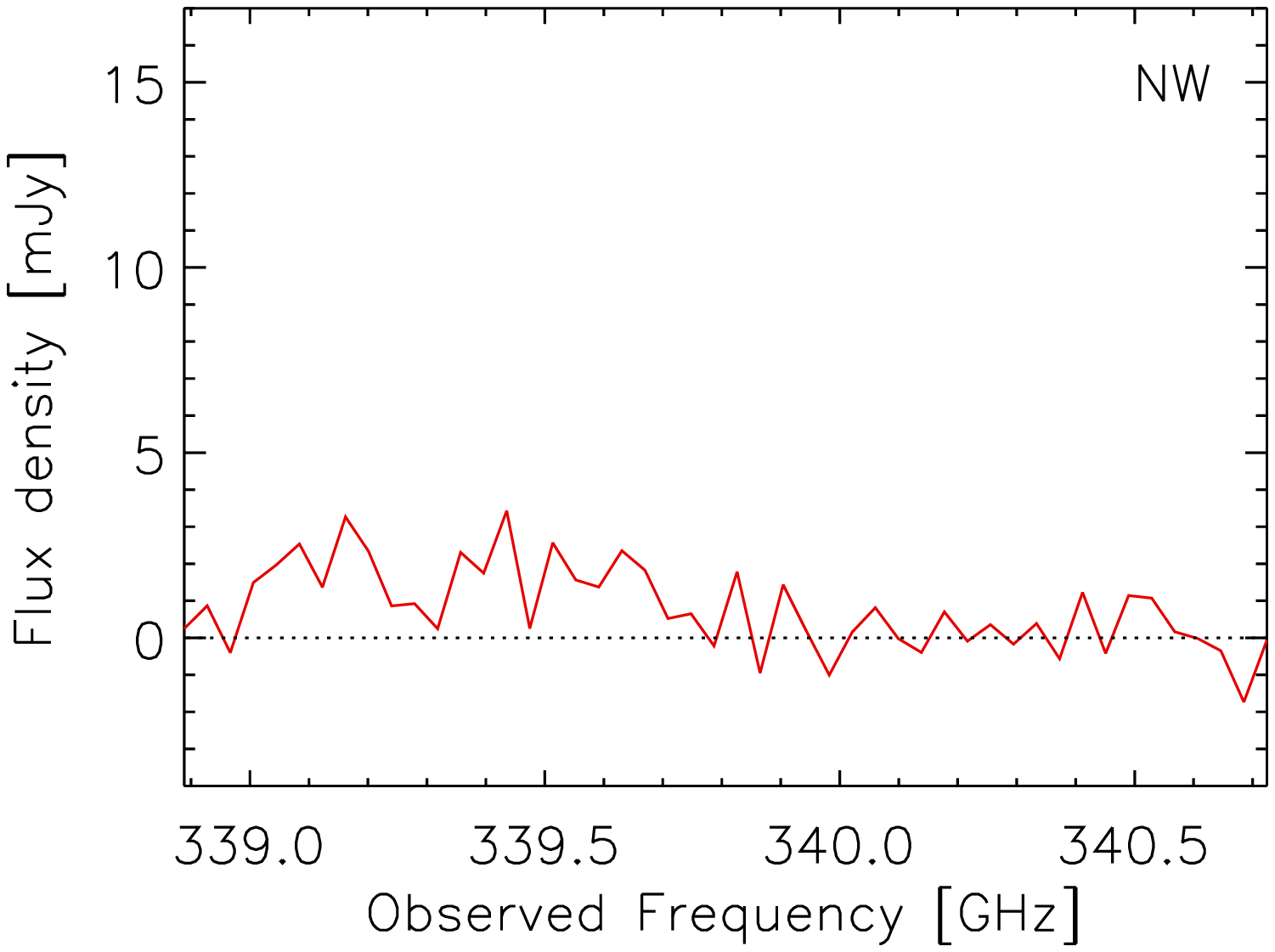}\plotone{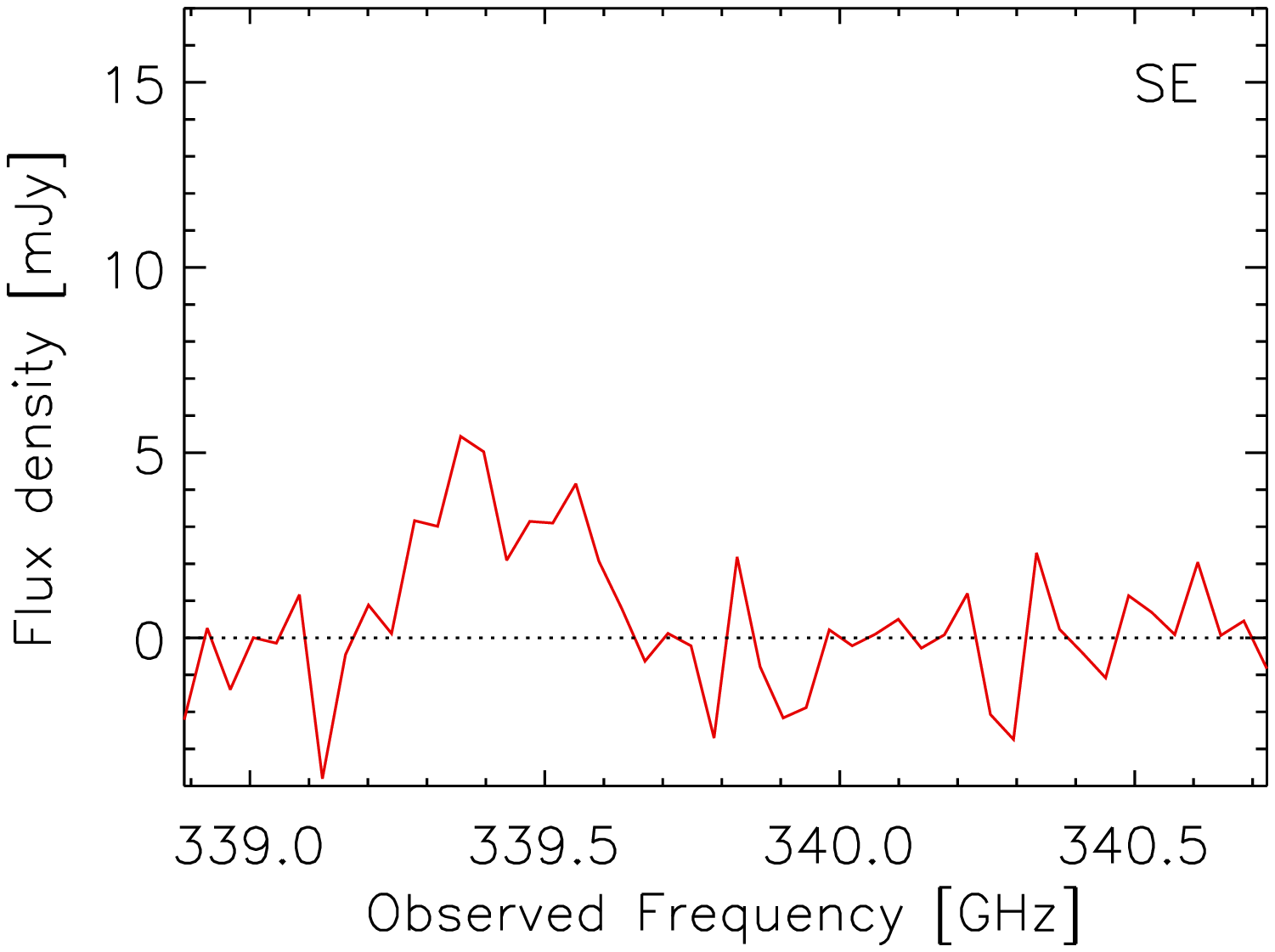}
\vspace{.25cm}
\caption{\footnotesize Left: Continuum-subtracted spectral profile of the \CIIno\, line extracted from the ALMA image using a 1\arcsec-diameter circular aperture centered on W2246-0526 (red solid line), with a Gaussian fit shown by the dashed blue line. The red wing of the line lies partially outside of the observed spectral window (see text for details). Middle and right: Spectra of the NW and SE components, respectively. The spectrum of the NW component was extracted with a rectangular aperture of 0.75\arcsec\,$\times$\,0.5\arcsec\, and that of the SE component with a circular aperture of diameter 1\arcsec.}
\label{f:spectra}
\vspace{.25cm}
\end{figure*}

\begin{figure}[t!]
\epsscale{0.94}
\plotone{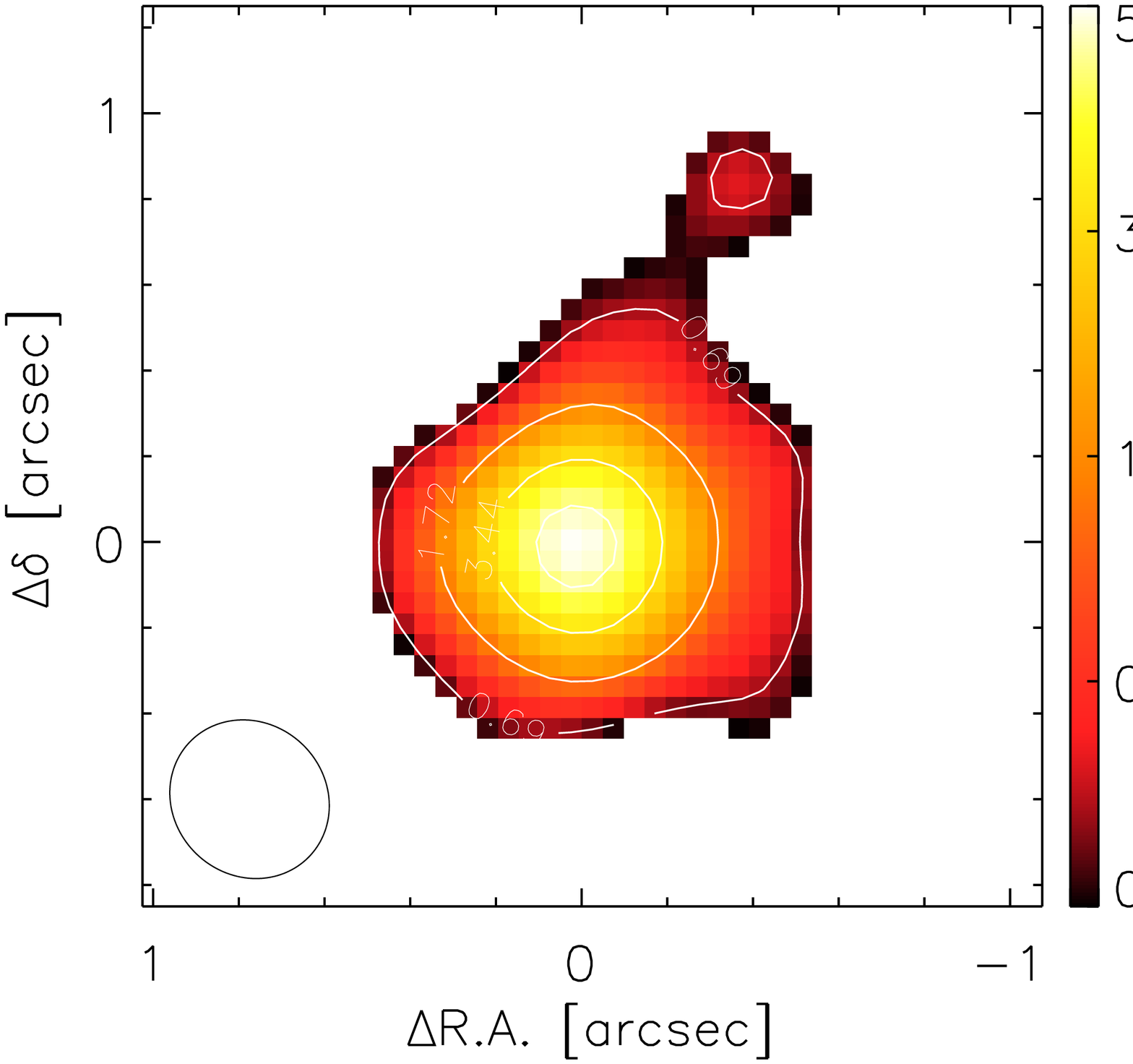}
\vspace{1.cm}
\plotone{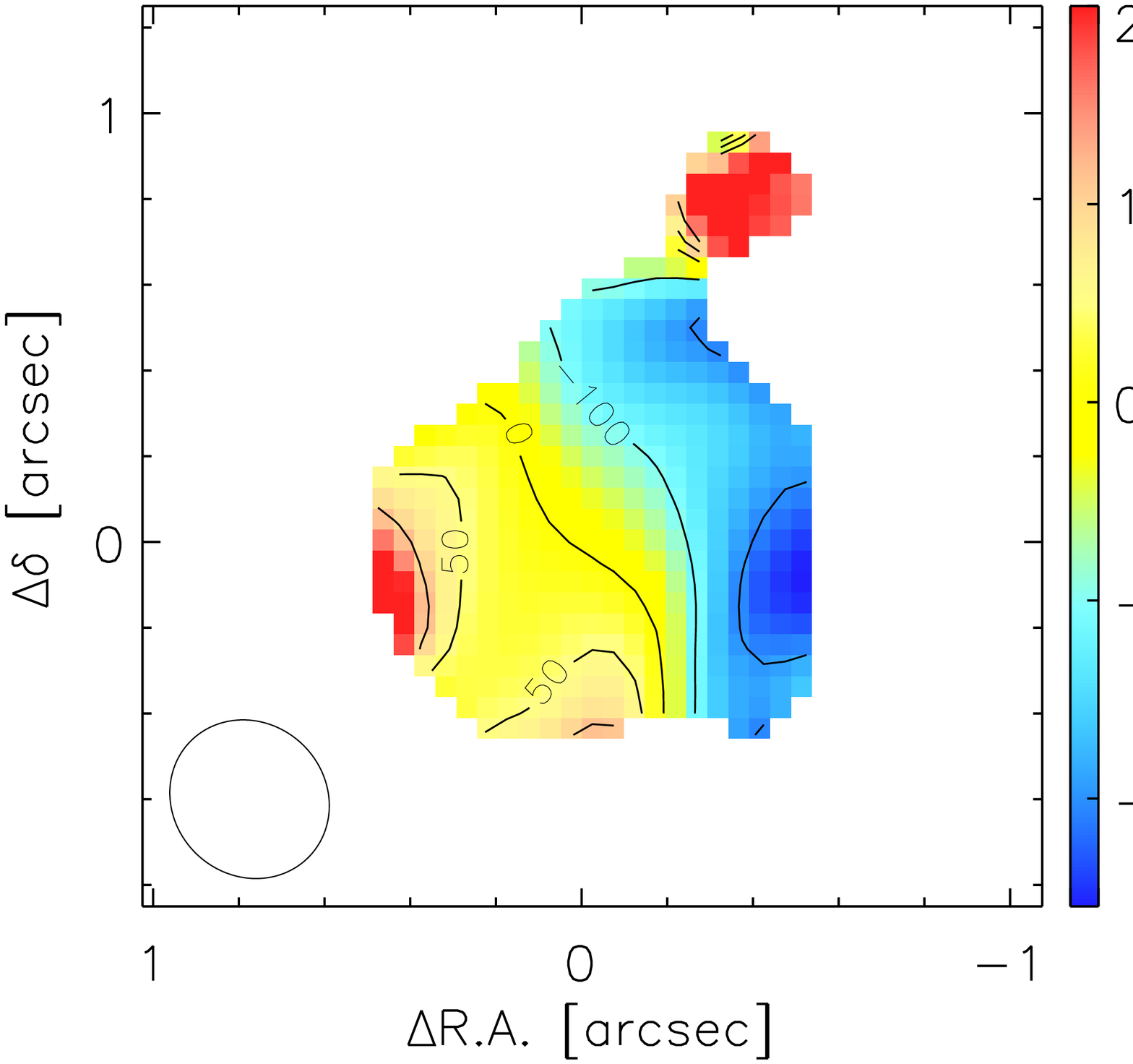}
\vspace{1.cm}
\plotone{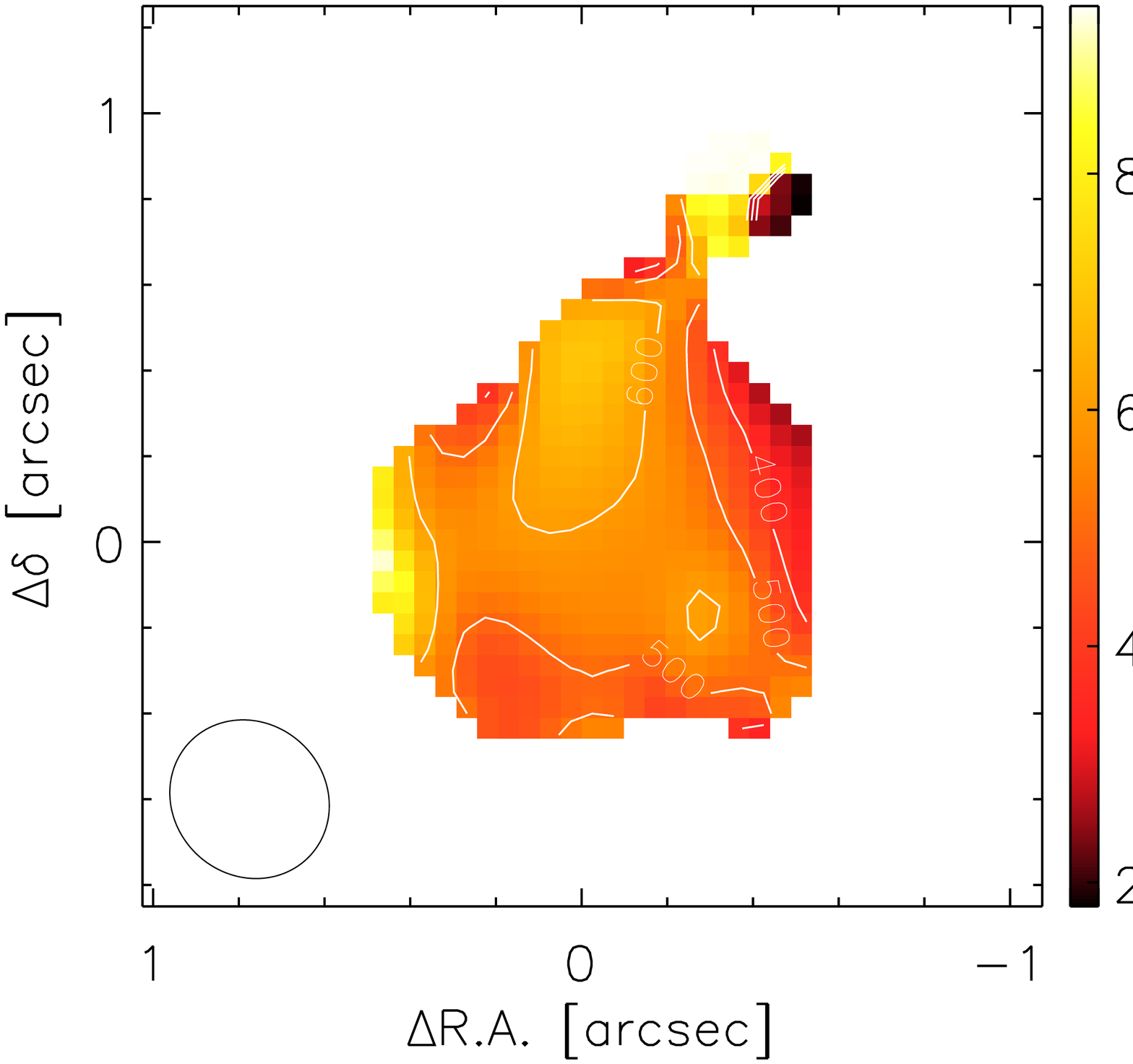}
\vspace{1.cm}
\caption{\footnotesize Moment maps of the central region of W2246-0526 where the \CIIno\, S/N$\,\geq\,$3: total intensity (top), projected velocity field (middle), and velocity dispersion (bottom). The contours in the panels are: [4,10,20,30]$\,\times\,\sigma$, [--200,--100,0,50,100]\,\kms, and [400,500,600]\,\kms, respectively. The clean beam is shown at the bottom-left corner of the images. The range of the projected radial velocity across the galaxy is modest, $\sim\,$150\,\kms; a few “hot-spots” have velocities up to +/– 200 \kms\, with respect to the dynamical center of the system, defined by the \CIIno\, peak. The resolved velocity dispersion is remarkably large and uniform across the whole system, with FWHM $\sim\,500-600$\,\kms\, even at kpc distances from the SMBH.
}\label{f:moments}
\vspace{.25cm}
\end{figure}

Our ALMA data reveal strong, spatially-resolved emission in both \CIIno\, and dust continuum. W2246-0526 is also detected in \textit{Hubble Space Telescope} (\textit{HST}) WFC3 rest-frame UV images (F160W; see Figure~\ref{f:images}), which reveal emission from unobscured, young, massive stars and the AGN. Furthermore, we detect \CIIno\, and UV emission from two additional, less luminous galaxies located 6.6 kpc northwest and 33\,kpc southeast of W2246-0526, marked NW and SE, respectively, in the top panel of Figure~\ref{f:images}. The spectra of W2246-0526 and its two companions are presented in Figure~\ref{f:spectra}. The moment maps are shown in Figure~\ref{f:moments} and will be discussed in section~\ref{s:discussion}. The ALMA and \textit{HST} morphologies support the idea that gravitational lensing is not important in Hot DOGs \citep{Tsai2015,Wu2014}. This is in strong contrast to many other well-studied dusty, IR-bright, high-redshift systems \citep[e.g.,][]{Vieira2013}.

Considering the contribution from the ALMA beam, the intrinsic size (FWHM) of the resolved \CIIno\, emission is 2.5\,$\pm$\,0.3\,kpc, about the same as in the ultraviolet \textit{HST} image (see Figure~\ref{f:images}), which suggests that part of the \CIIno\, emission could be associated to extended star formation in the host galaxy, not the AGN. In contrast, the continuum emission is only 1.3\,$\pm$\,0.5\,kpc wide, suggesting that it could be in part contributed by the AGN. This difference between ionized gas and dust emission sizes is often seen in very high-redshift quasars \citep{Wang2013} and star-forming galaxies \citep{Capak2015}, although it is not ubiquitous \citep{Kimball2015}. In star-bursting galaxies most of the \CIIno\, emission arises from dense PDRs close to massive stars, with a possible important contribution from low-density, warm ionized gas heated by the dissipation of turbulent energy in shocks \citep{Appleton2013}. In addition, although AGN emission is not expected to power \CIIno\, \citep{Langer2015}, AGN-driven outflows are sometimes able to periodically inject enough energy into the ISM to enhance the \CIIno\, emission and drastically affect the kinematics of the gas in the host galaxy \citep{Guillard2015} (see section~\ref{ss:blowing}).

\begin{figure}[t!]
\epsscale{0.94}
\plotone{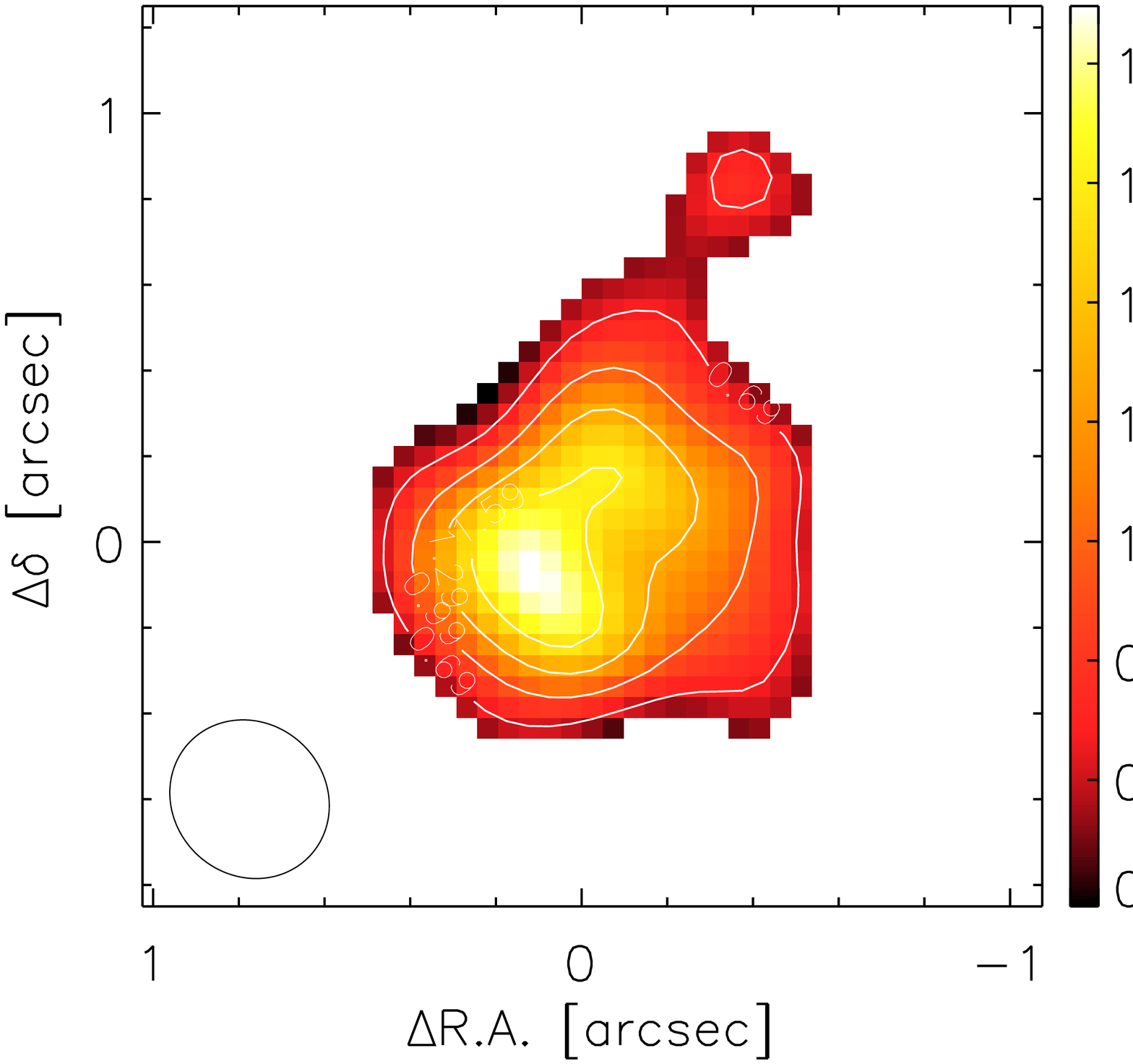}
\vspace{1.cm}
\plotone{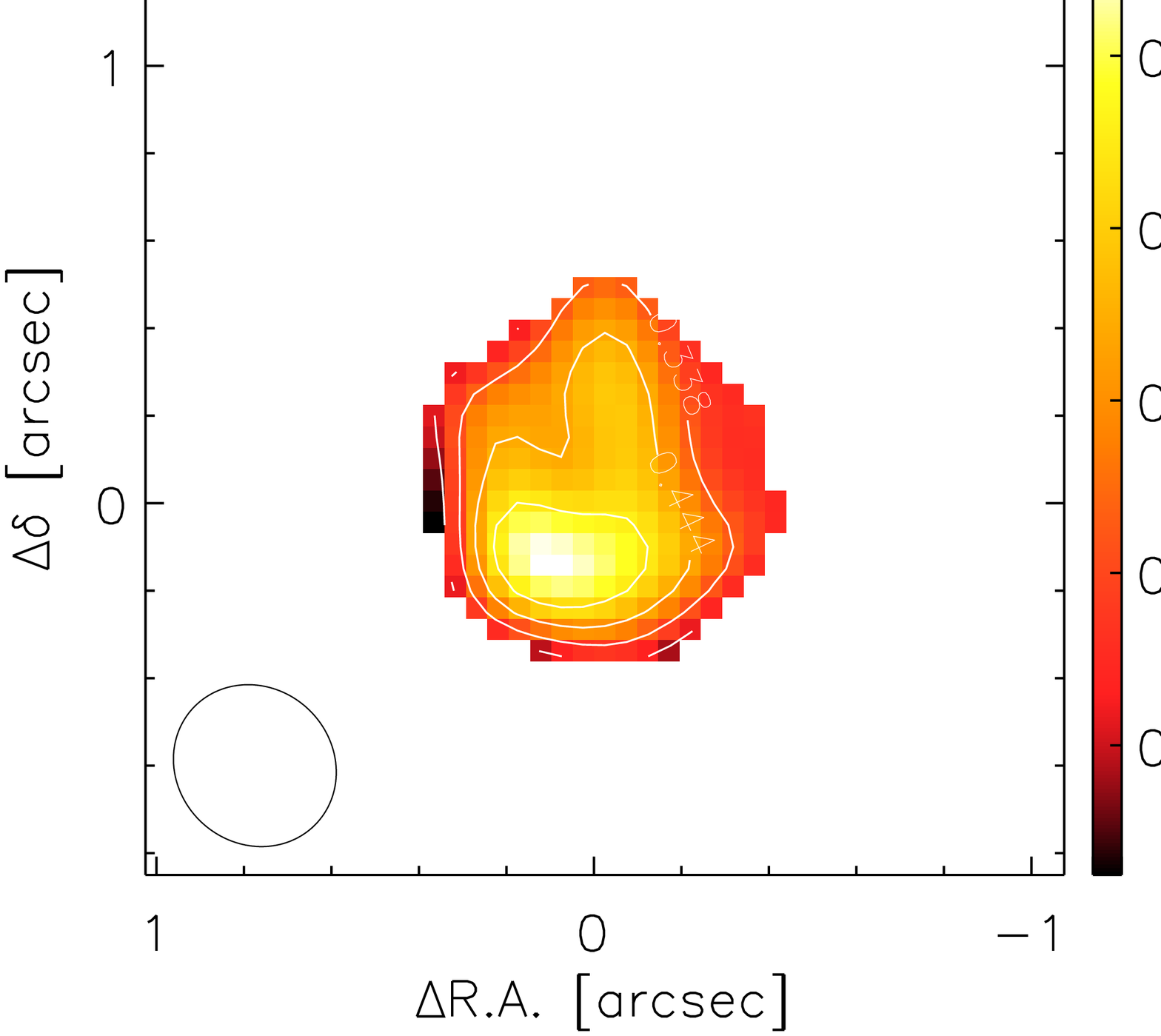}
\vspace{1.cm}
\plotone{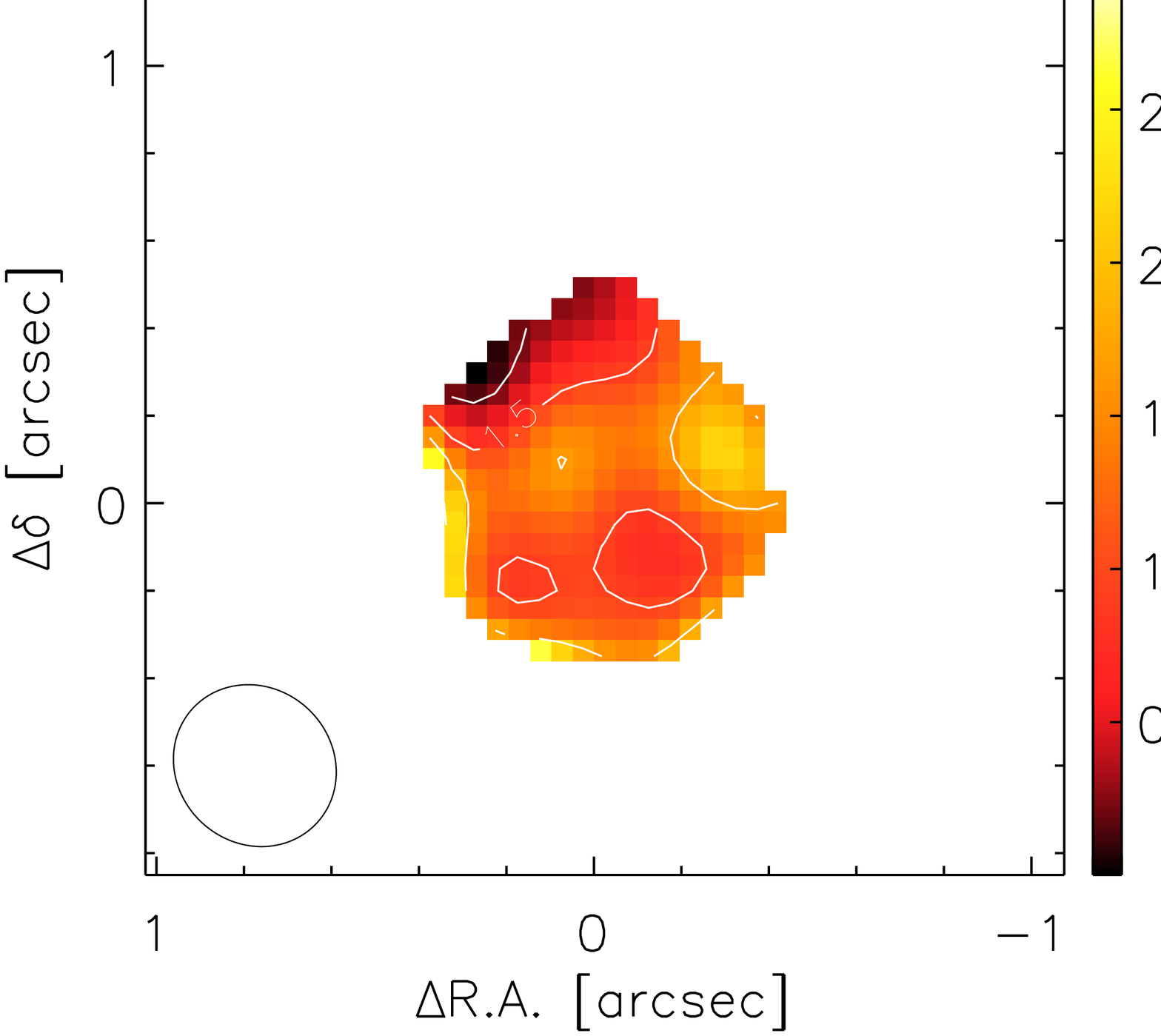}
\vspace{1.cm}
\caption{\footnotesize Residual underlying \CIIno\, line (top) and dust continuum emission (middle) in W2246-0526 after subtracting a central point source (see section~\ref{s:results}), where emission is detected at S/N$\,\geq\,$3. The clean beam is shown at the bottom-left corner of the images. Both images display extended structure well aligned with the velocity gradient seen in Figure~\ref{f:moments} (middle). The bottom panel shows the rest-frame \CIIno\, equivalent width in the underlying host galaxy, as measured using the point-source-subtracted images. Values range between $\sim\,$1--2\,$\mu$m.}
\label{f:psfsub}
\vspace{.25cm}
\end{figure}

%%AWB - QUESTION - Do we want the pre-subtracted ALMA images (Fig 5) to come here ahead of Fig 3? 

In order to study the underlying emission from the host galaxy in W2246-0526, we used the clean beam of the observations to subtract an unresolved central component from both the \CIIno\, and dust continuum images. This provides the first insight into the spatial distribution of ionized gas in the host galaxy of an hyper-luminous AGN (see Figure~\ref{f:psfsub}). The beam was scaled to minimize the dispersion in the central 0.5\arcsec\, region of the residual map, which required scaling factors of 79\% and 89\% of the peak flux in the line and continuum, respectively. Within a 2\arcsec-diameter aperture, the subtracted images retain 67\% and 28\% of the total luminosity in \CIIno\, and the continuum respectively. Thus, most of the \CIIno, and around a quarter of the dust continuum emission originates several kpc away from the AGN. The residual extended emission structure is oriented consistently with the mild velocity gradient discussed in section~\ref{ss:dispersion}, and with a surface brightness that varies by as much as a factor of three across the galaxy. Figure~\ref{f:psfsub} also shows a resolved map of the rest-frame \CIIno\, equivalent width from the underlying host galaxy of W2246-0526. The values are very uniform, between $\sim\,$1--2\,$\mu$m, similar to those found in local IR-luminous galaxies \citep{DS2013}, and consistent with the host being an actively star-forming galaxy.

\section{Discussion}\label{s:discussion}

\subsection{The \CIIno/FIR\, Deficit}\label{ss:deficit}

The \CIIno/FIR\, emission ratio provides insight into the physical state of the ISM in galaxies. In star-forming galaxies, this ratio is sensitive to the average intensity of the ionizing radiation field \citep{Kaufman1999}, with pure star-forming systems having values an order of magnitude greater than AGN hosts \citep{DS2013}. The \CIIno/FIR\, ratio is reduced in the nuclei of galaxies, in compact starbursts, and in general in regions with greater luminosity surface densities, regardless of the presecence of an AGN \citep{DS2014}. However, note that all galaxies harboring deeply obscured AGN, like W2246-0526, show \CIIno/FIR\, ratios less than $\sim\,10^{-3}$ \citep{Stacey2010,Wagg2010,DS2013,Carilli2013}. 

\begin{figure}[t!]
\vspace{.25cm}
\epsscale{1.15}
\plotone{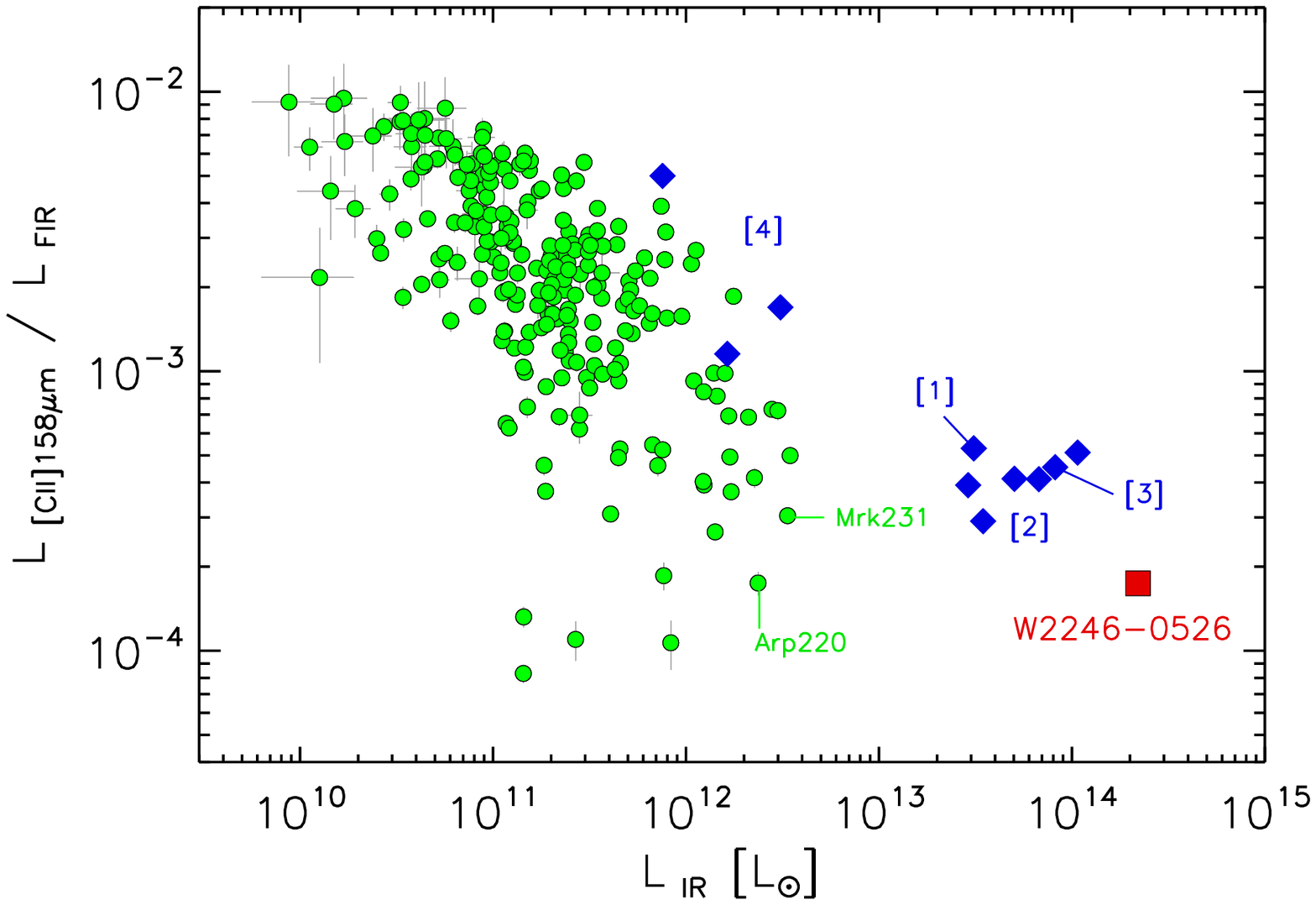}
\vspace{.25cm}
\caption{\footnotesize The \CIIno-to-FIR$_{[42-122\,\mu {\rm m}]}$ luminosity ratio as a function of the \LIR$_{[8-1000\,\mu {\rm m}]}$ of galaxies. The green circles represent nuclear values from the 60\,$\mu$m-selected, flux-limited sample of local luminous IR galaxies from GOALS \citep{Armus2009,DS2013}. W2246-0526, shown as a red square, has one of the largest line-to-FIR deficits seen at high-redshift, and can only be compared to the most obscured, compact systems observed in the local Universe today. For reference, we also show high-redshift ($z\,>\,4$) quasars as blue diamonds, with the associated reference next to the data-points: [1] \cite{Wagg2010}, group [2] \cite{Wang2013}, [3] \cite{Kimball2015}, group [4] \cite{Willott2013,Willott2015}. For quasars without an estimate of the total IR luminosity, we have conservatively assumed $\LIR_{[8-1000\,\mu {\rm m}]}$\,=\,6.3$\,\times\,\LFIR_{[42-122\,\mu {\rm m}]}$, the ratio measured in W2246-0526. For quasars with only $L_{\rm bol}$ available we have assumed $\LIR$\,=\,0.63\,$\times$\,$L_{\rm bol}$, also as in W2246-0526.
%For high-z star-forming systems \citep{DeBreuck2014,Riechers2013} we have assumed $\LIR_{[8-1000\,\mu {\rm m}]}$\,=\,1.8$\,\times\,\LFIR_{[42-122\,\mu {\rm m}]}$, similar to their local IR-luminous counterparts \citep{Armus2009}.
}\label{f:CIIdeficit}
\vspace{.25cm}
\end{figure}

The total \CIIno\, luminosity of W2246-0526 is $L_{\rm [C\,{\sc II}]}$\,=\,6.1\,($\pm$\,0.4)\,$\times\,10^9\,\Lsun$, and a multi-wavelength fit to its integrated SED yields a conservative 
estimate of its IR luminosities: $\LIR_{[8-1000\,\mu {\rm m}]}$\,=\,2.2 ($\pm\,$0.4)$\,\times\,10^{14}\,\Lsun$, and $\LFIR_{[42-122\,\mu {\rm m}]}$\,=\,3.5\,($\pm$\,0.4)$\,\times\,10^{13}\,\Lsun$\footnote{These luminosities were obtained by interpolating between the photometric data points using power laws and integrating under the curve. The resulting values are thus likely slightly underestimated. See \cite{Eisenhardt2012} and \cite{Tsai2015} for more details.}. This results in an integrated \CIIno/FIR\, ratio of $\sim\,2\,\times\,10^{-4}$ (see Figure~\ref{f:CIIdeficit}), which is likely an upper limit to the nuclear value since the \CIIno\, emission is twice as extended as the dust continuum (section~\ref{s:results}). Nevertheless, this ratio is still only matched in the most obscured nuclei of local dusty IR-luminous systems, like Mrk~231, the classical nearby archetype of a dust-enshrouded AGN. Yet, the \CIIno\, luminosity of W2246-0526 is more than one order of magnitude greater than that of Mrk~231 and the velocity dispersion is $>\,$2 times larger \citep{Fischer2010}, implying a much more massive and turbulent environment. Using the resolved, rest-frame 157.7$\mu$m continuum emission (section~\ref{s:obs}) as a measure of the starburst size, the IR luminosity surface brightness of W2246-0526 is $\Sigma_{\rm IR}\,=\,1.7\,\times\,10^{14}\,\Lsun$, which is in excellent agreement with the extrapolation of the correlation between the \CIIno/FIR\, ratio and $\Sigma_{\rm IR}$ found for local luminous IR galaxies \citep{DS2013,DS2014}.

\subsection{The Strikingly Uniform Velocity Disperision of W2246-0526}\label{ss:dispersion}

Figure~\ref{f:moments} present the moment maps of the \CIIno\, line emission in W2246-0526. The complex iso-velocity contours (middle panel) indicate that the galaxy is neither dynamically relaxed nor is there a cool disk present. There is a strikingly uniform, broad velocity dispersion with FWHM $\simeq\,500-600$\,\kms\, ($\sigma\,\sim\,210-250$\,\kms) across the entire \CIIno\, emitting region, coupled with a modest shear, consistent with bulk rotation on several-kpc scales with a maximum projected line-of-sight velocity of just $\sim\,$150\,\kms\, (Figure~\ref{f:moments}, bottom and middle panels respectively). In fact, following \cite{DeBreuck2014}, we calculated that rotational shear can contaminate the velocity dispersion by only $<\,10$\%. A few spots show peak radial velocities of up to $\pm$\,200\,\kms\, with respect to the dynamical center of the system, indicating that the host may be undergoing a minor merger \citep[similar to][]{Riechers2014}, perhaps with the NW source located 6.6 kpc away (see Figure~\ref{f:images}).

The extent and velocity structure of the \CIIno\, emission in W2246-0526 leads to two very important points. First, the high velocity dispersion gas extends throughout the entire resolved system, and is not solely associated with a hypothetical central stellar bulge around the AGN; second, the high dispersion-to-shear ratio suggests an extremely turbulent ISM. This is in contrast with most high-redshift star-forming galaxies and quasars detected by ALMA in \CIIno, which show kinematic signatures consistent with rotation-supported disks \citep{Wang2013,DeBreuck2014,Kimball2015}, or highly irregular velocity dispersion fields \citep{Riechers2013} with morphologies typical of major mergers, like those seen in sub-millimeter galaxies at $z\,\sim\,2$ \citep{Tacconi2008,Engel2010}. A velocity dispersion as great as that of W2246-0526 has only been found in a few nearby systems. The best prototype is probably the interacting starburst galaxy NGC~6240, which hosts two deeply buried SMBHs in its core. \textit{Herschel Space Observatory} FIR spectra of this galaxy show resolved \CIIno\, line emission with a velocity dispersion of 600\,$\pm$\,65\,\kms. 
%%AWB - QUESTION - Not sure the CO is relevant here, since we have CII directly in NGC6240, and no CO in 2246? Delete next sentence? 
%Low-velocity shock excitation is able to explain at least the CO kinematic properties in NGC~6240 \citep{Meijerink2013}.

\subsection{An Isotropic, Large-Scale Blow Out of the Gas}\label{ss:blowing}

The observed kinematic properties of the \CIIno\, line agree with quantitative arguments regarding the stability of such a luminous system in terms of the energy and momentum injected by the AGN into the ISM. For a given velocity dispersion, $\sigma$, a system with a luminosity greater than: 

\begin{equation}
L_{\rm M}\,=\,\bigg(\frac{4 f_{\rm g} c}{G}\bigg) \sigma^4 \simeq\, 7.8\,\times\,10^{12}\,f_{\rm g 0.1}\,\sigma^4_{200}\,\,\Lsun
\end{equation}

\noindent
cannot be stable in terms of the radiation pressure of photons on dust grains alone \citep{Murray2005,King2011}; $f_{\rm g}$ is the gas fraction, with $f_{\rm g 0.1}$\,=\,$f_{\rm g}$/0.1 and $\sigma_{200}$\,=\,$\sigma$/200\,\kms. $L_{\rm M}$ can be understood here as an Eddington limit for an isotropic, momentum-driven wind. Even for an unrealistic extreme $f_{\rm g}\,\approx\,$1, the bolometric luminosity of W2246-0526 is at least twice the derived threshold ($\sim\,$1.5$\,\times\,10^{14}\,\Lsun$), suggesting that the AGN is likely blowing away all the gas from the galaxy. Furthermore, in this scenario, the ejection of the gas would not be bipolar (like in M82) but rather isotropic \citep{Murray2005}. Indeed, the striking uniformity of the high velocity dispersion resolved right across the galaxy strongly suggests that turbulent gas is expanding in all directions.

A similar argument can be made in terms of the total energy that is being injected into the ISM to create an expanding isothermal bubble through an energy deposition-driven wind \citep{Murray2005,King2011}. Here the limiting luminosity is given by:

\begin{equation}
L_{\rm E}\,\sim\,\bigg(\frac{100}{\xi}\bigg)\bigg( \frac{4 f_{\rm g} c}{G}\bigg)\,\sigma^5\,\simeq\,\,5.2\,\times\,10^{12}\,f_{\rm g 0.1}\,\xi^{-1}_{0.1} \sigma^5_{200}\,\,\Lsun
\end{equation}

\noindent
where $\xi$ is the efficiency of energy transfer into the ISM, with $\xi_{0.1}$\,=\,$\xi$/0.1. W2246-0526 also exceeds this stability threshold, even for high gas fractions and low efficiencies. This matches observations of high-velocity, energy-driven molecular outflows in local galaxy mergers \citep{Veilleux2013,Tombesi2015}, but in the case of W2246-0526 at much larger scales and with a higher degree of isotropy, as suggested by the observed uniformity of its kpc-scale velocity dispersion.

\section{Summary}

The \CIIno\, emission from W2246-0526 has a large, uniform velocity dispersion and low shear. The system is unstable in terms of both the energy and momentum injected by the accreting SMBH into the ISM. This strongly argues that W2246-0526 is blowing out its ISM isotropically, in an unprecedented, homogeneous, large-scale turbulent outflow with a greater covering factor than the winds seen in high-redshift QSOs and local ultra-luminous IR galaxies. This suggests that W2246-0256 is near to bursting out of its dusty cocoon to become a powerful optically visible QSO. Caught at a time when the Universe was ramping up to its peak of star formation and SMBH accretion, our ALMA observations clearly reveal extreme conditions in the ISM of the most luminous galaxy known, where the feedback from the powerful AGN is having a strong impact on the evolution and fate of the entire galaxy \citep{Silk1998}.

\section*{Acknowledgments}

The authors would like to thank H. Jun for suggestions provided during the analysis of the data. T.D-S. acknowledges support from ALMA-CONYCIT project 31130005 and FONDECYT 1151239. R.J.A. acknowledges support from Gemini-CONYCIT project 32120009 and FONDECYT 1151408. The work of C-W.T., P.E., D.S. and C.B. was carried out at the Jet Propulsion Laboratory, California Institute of Technology, under a contract with NASA. M.A. acknowledges partial support from FONDECYT through grant 1140099. ALMA is a partnership of ESO (representing its member states), NSF (USA) and NINS (Japan), together with NRC (Canada) and NSC and ASIAA (Taiwan), in cooperation with the Republic of Chile. The Joint ALMA Observatory is operated by ESO, AUI/NRAO and NAOJ.
%This work is based in part on observations made with the \textit{Spitzer Space Telescope}, the \textit{Herschel Space Observatory}, the \textit{Wide-field Infrared Survey Explorer}, as well as archival data from the NASA/ESA \textit{Hubble Space Telescope}.\\

%\bibliographystyle{/Users/tanio/mypapers/apj}
%\bibliography{/Users/tanio/mypapers/bib}{}

\end{document}